\documentclass{moriond}

\usepackage{fancyhdr}
\newcommand{\Photo}{}

\newcommand{\ttbar}{$\mathrm{t}\bar{\mathrm{t}}$}
\newcommand{\fbinv}{\,fb\textsuperscript{$-1$}}
\newcommand{\pbinv}{\,pb\textsuperscript{$-1$}}
\newcommand{\pt}{$p_{\mathrm{T}}$}
\newcommand{\abseta}{$|\eta|$}

\begin{document}
\vspace*{4cm}
\title{FIRST RESULTS OF THE CMS EXPERIMENT AT 13.6\,TeV}

\author{J. KNOLLE~\footnote{\url{joscha.knolle@cern.ch}}}

\address{Universiteit Gent, Vakgroep Fysica en Sterrenkunde, Proeftuinstraat 86,\\ 9000 Gent, Belgium\\[0.7\baselineskip]\textnormal{on behalf of the CMS Collaboration}}

\maketitle\abstracts{%
    The CMS experiment recorded about 38.5\fbinv\ of proton-proton collision data at \mbox{$\sqrt{s}=13.6$\,TeV} in 2022, the first year of the LHC Run~3.
    Performance highlights on electron triggers, tracking with muons, and jet energy calibration are shown.
    The CMS luminosity measurement and an independent cross-check based on the counting of Z~boson events are discussed.
    Finally, the first Run~3 cross section measurement of top quark pair production is presented.
}

\section{Introduction}

The CERN LHC resumed data taking in 2022 with proton-proton collisions at the unprecedented center-of-mass energy of 13.6\,TeV, marking the start of its Run~3.
The CMS experiment~\cite{CMS:Detector-2008} measured a total delivered luminosity of about 42.0\fbinv.
Data was recorded with an efficiency of 92\%, resulting in a recorded data set of about 38.5\fbinv.
89\% of the recorded data are certified as ``good for physics'', meaning that all CMS subdetectors were fully operational.
In this contribution, I present first results of the CMS Collaboration with the new data.

\section{Performance highlights}

\begin{figure}[t!]
\centering
\raisebox{4pt}{\includegraphics[width=0.31\textwidth]{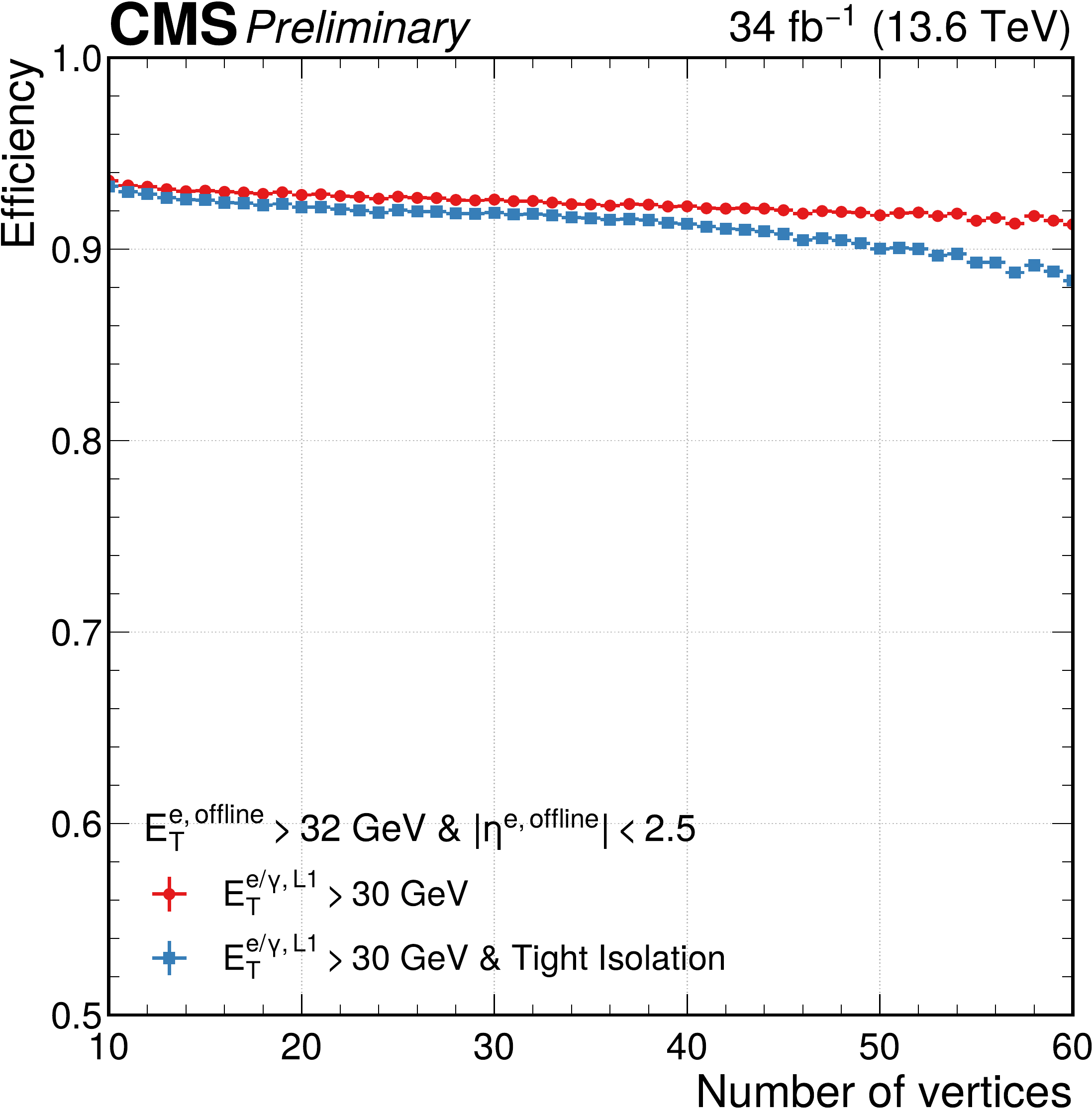}}%
\hfill%
\raisebox{-1pt}{\includegraphics[width=0.327\textwidth]{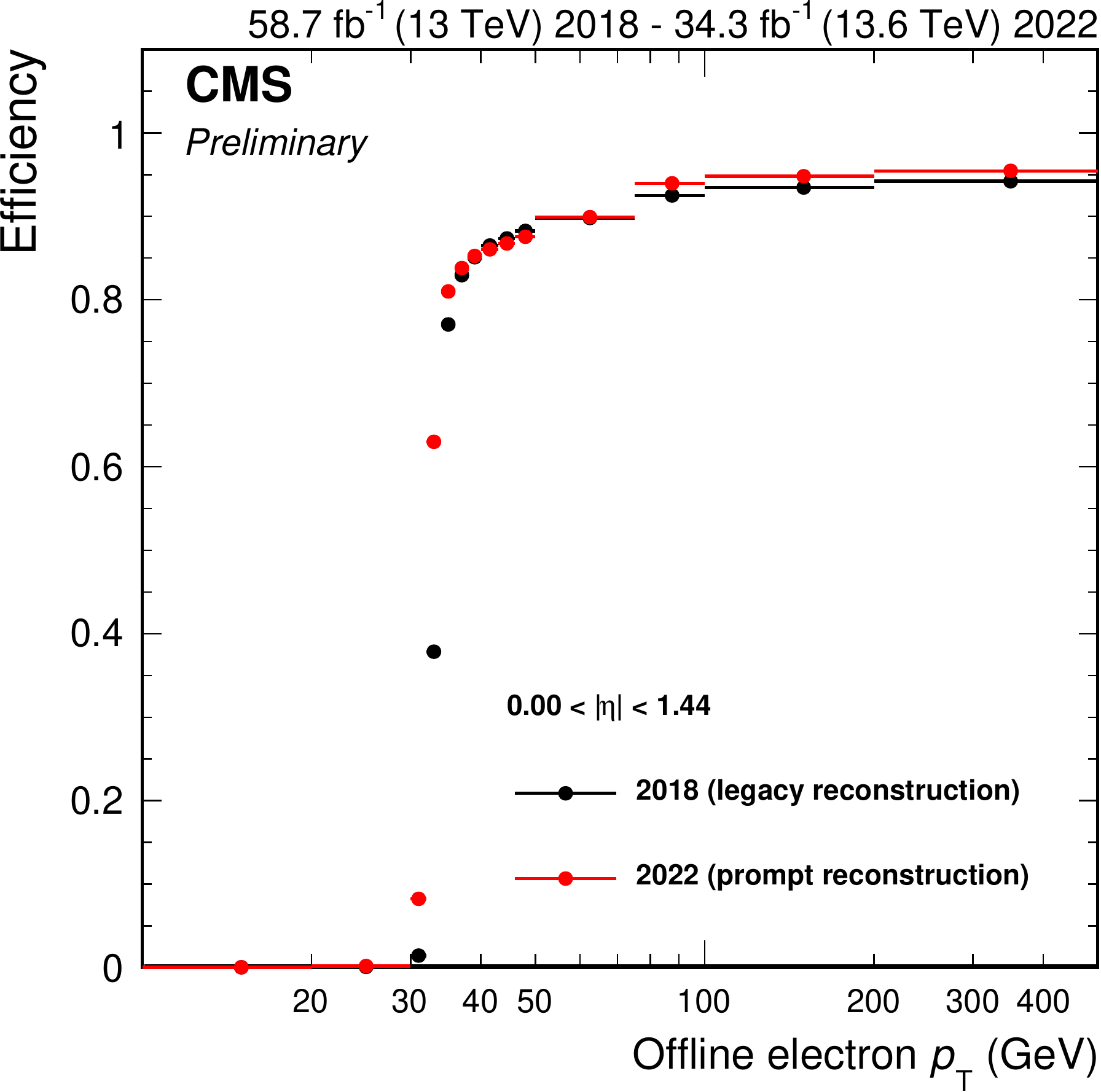}}%
\hfill%
\includegraphics[width=0.335\textwidth]{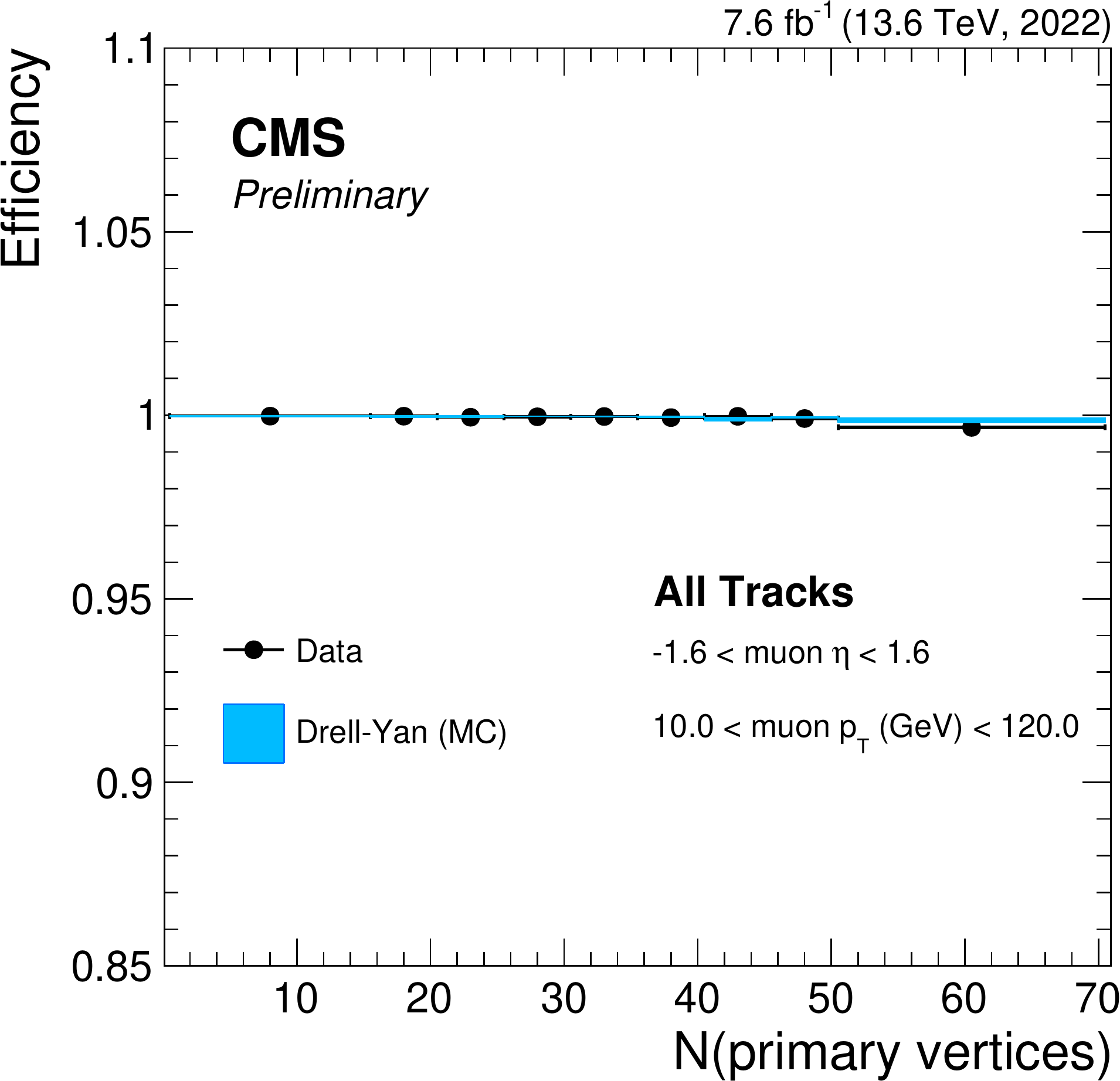}
\caption{%
    (Left) L1 trigger efficiency as function of the offline number of vertices for nonisolated (red) and isolated (blue) electron/photon candidates~\protect\cite{CMS:DP-2023-008}.
    (Center) HLT trigger efficiency as function of the offline electron \pt\ for the path with a \pt\ threshold of 32\,GeV in 2018 (black) and 2022 (red)~\protect\cite{CMS:DP-2023-015}.
    (Right) Tracking efficiency as function of the number of vertices for data (black points) and simulation (blue)~\protect\cite{CMS:DP-2022-046}.
}
\label{fig:electronmuon}
\end{figure}

A two-tiered trigger system is used to select events of interest, with the level-1 trigger (L1) selecting events at a rate of around 100\,kHz within a fixed latency of about 4\,\textmu s~\cite{CMS:TRG-17-001}, and the high-level trigger (HLT) reducing the event rate to around 1\,kHz before data storage~\cite{CMS:TRG-12-001}.
L1 electron/photon candidates are reconstructed by clustering calorimeter trigger towers around a seed with local transverse energy $E_{\mathrm{T}}>2\,\mathrm{GeV}$.
Identification and isolation requirements are imposed on the candidate clusters and calibrated from look-up-tables.
At the HLT, a version of the full event reconstruction software optimized for fast processing is run, and electron candidates are subject to more complex identification and isolation requirements.
The single-electron HLT path with a \pt\ threshold of 32\,GeV is vastly used in physics analyses at CMS.
The performance of the L1 and HLT electron reconstruction is evaluated with the tag-and-probe method~\cite{CMS:EWK-10-002} using $\mathrm{Z}\rightarrow\mathrm{e}^+\mathrm{e}^-$ events from the full 2022 data set.
Results are shown in Fig.~\ref{fig:electronmuon} (left and center).
A high trigger efficiency with a good stability against pileup is found~\cite{CMS:DP-2023-008,CMS:DP-2023-015}.

Muons are reconstructed in the range $|\eta|<2.4$ by combining information from the tracker, the muon spectrometers, and the calorimeters in a global fit~\cite{CMS:MUO-16-001}.
The tag-and-probe method can be used with $\mathrm{Z}\rightarrow$\textmu$^+$\textmu$^-$ events to evaluate the efficiency of the track reconstruction~\cite{CMS:TRK-11-001} in the tracker.
Data recorded until 23 Aug 2022 is used, and a very good tracking efficiency over the full pileup range is found~\cite{CMS:DP-2022-046}.
Results are shown in Fig.~\ref{fig:electronmuon} (right).

\begin{figure}[t!]
\centering
\includegraphics[width=0.32\textwidth]{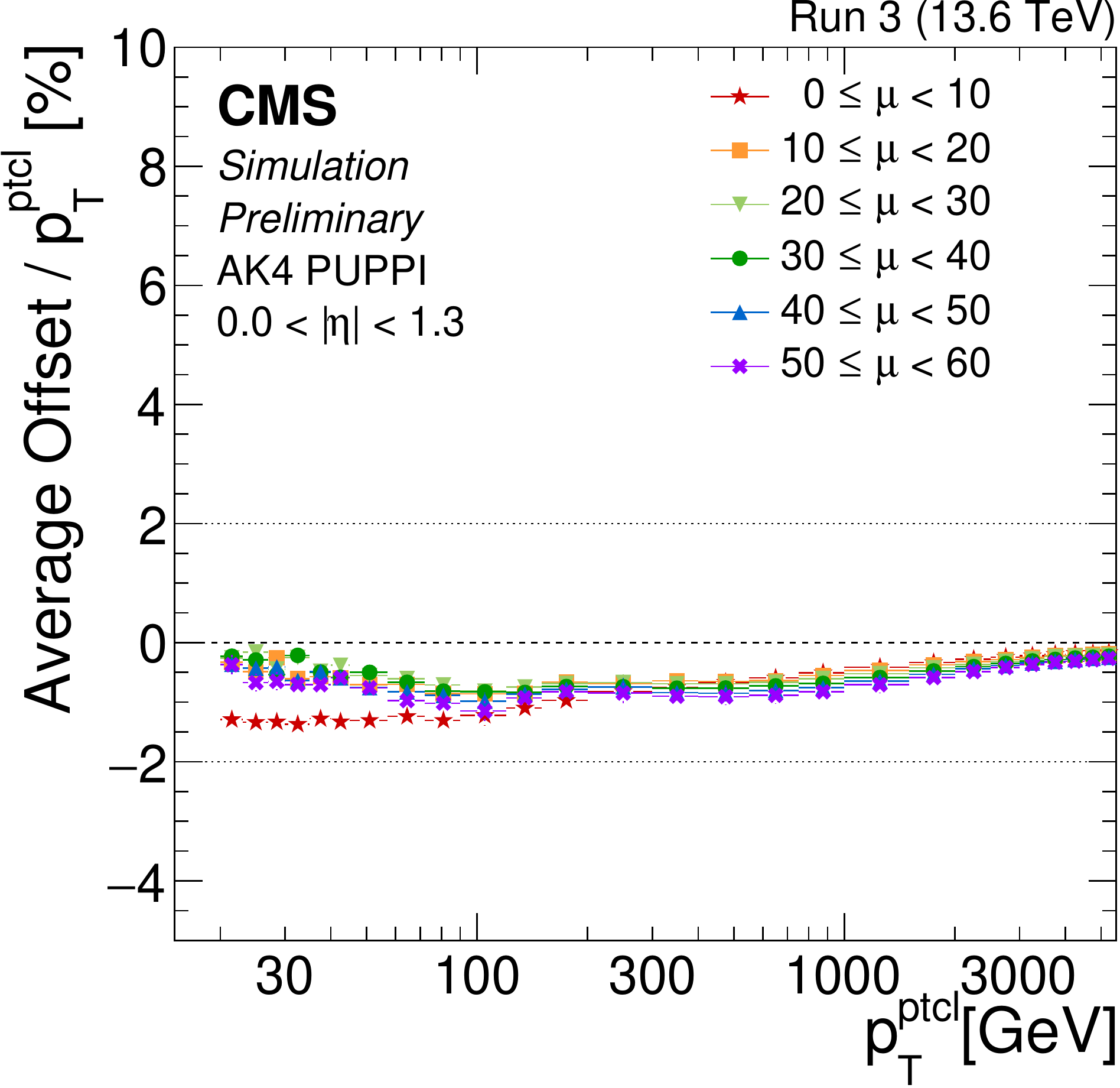}%
\hfill%
\raisebox{2pt}{\includegraphics[width=0.319\textwidth]{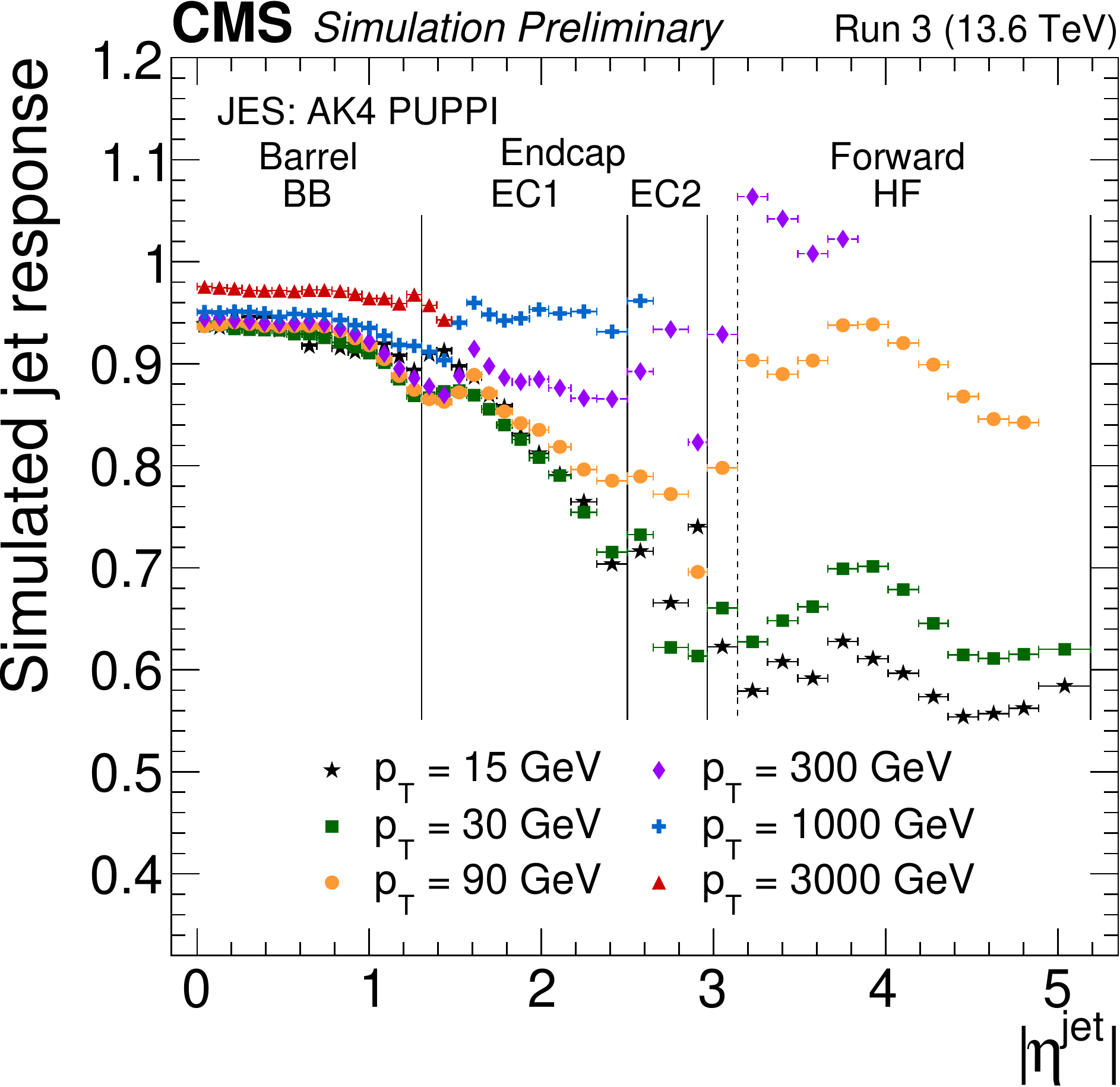}}%
\hfill%
\includegraphics[width=0.32\textwidth]{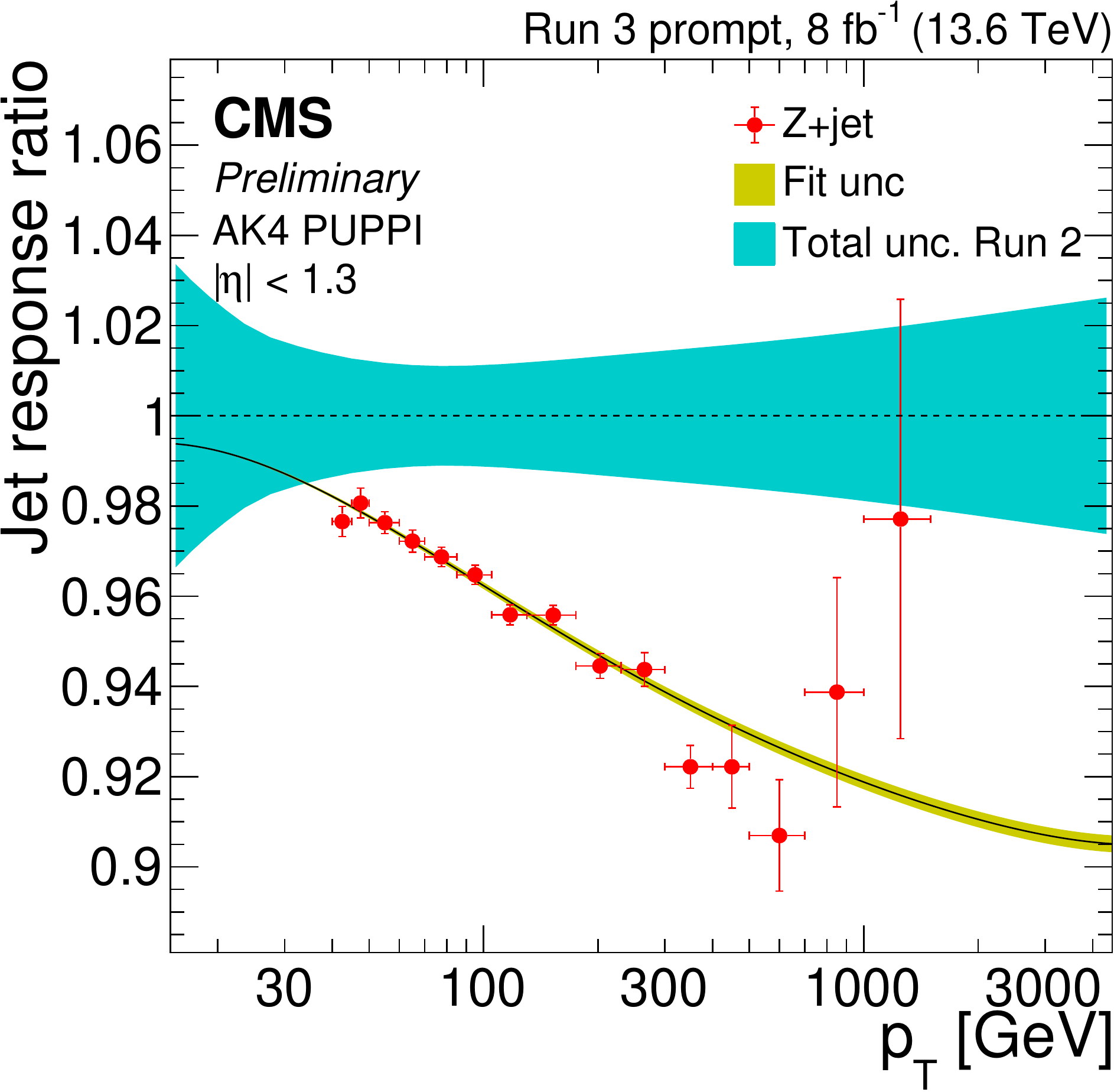}
\caption{%
    (Left) Pileup offset in simulation as function of \pt\ for different number of pileup interactions $mu$~\protect\cite{CMS:DP-2022-054}.
    (Center) Response correction from simulation as function of \abseta\ for different \pt\ values~\protect\cite{CMS:DP-2022-054}.
    (Right) Jet energy response measured in Z+jet events (red points) as function of \pt~\protect\cite{CMS:DP-2022-054}.
}
\label{fig:jets}
\end{figure}

Jets are clustered with the anti-$k_{\mathrm{T}}$ algorithm with a distance parameter of 0.4 (AK4).
The effect from pileup is mitigated with the pileup per particle identification algorithm (PUPPI)~\cite{Bertolini:Puppi-2014,CMS:JME-18-001}, which replaces charged hadron subtraction (CHS) as the main pileup mitigation algorithm in CMS for Run~3.
To calibrate the jet energy scale, corrections are derived from simulation to bring the measured jet response to that of particle-level jets on average.
Further measurements in standard candle processes are used to determine residual differences between data and simulation~\cite{CMS:JME-13-004}.
The first 2022 jet energy calibrations have been derived with the prompt reconstruction of the first 8\fbinv\ of data collected in 2022~\cite{CMS:DP-2022-054}, and results are shown in Fig.~\ref{fig:jets}.
The pileup offset in simulation (calculated as average \pt\ difference between jets in simulation with and without pileup) is much smaller than for CHS jets, and does not need to be applied as a correction.
The response correction from simulation shows a stable response in the barrel and a stronger \pt-dependence in the endcaps and forward calorimeter.
Some missing calorimeter calibrations in the prompt reconstruction lead to larger residual corrections.

\section{Measurement of the integrated luminosity}

\begin{figure}[t!]
\centering
\includegraphics[width=0.8\textwidth]{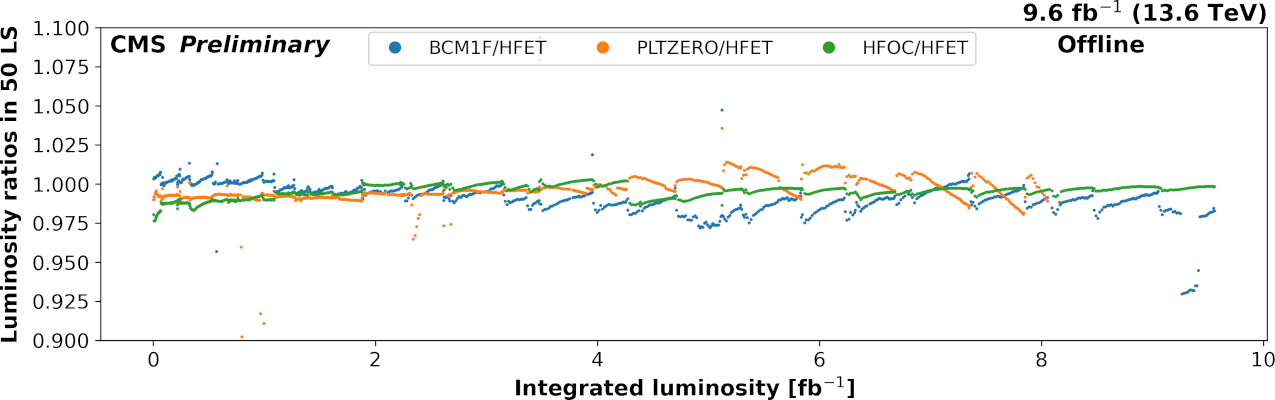} \\[5pt]
\hspace*{-0.008\textwidth}\includegraphics[width=0.804\textwidth]{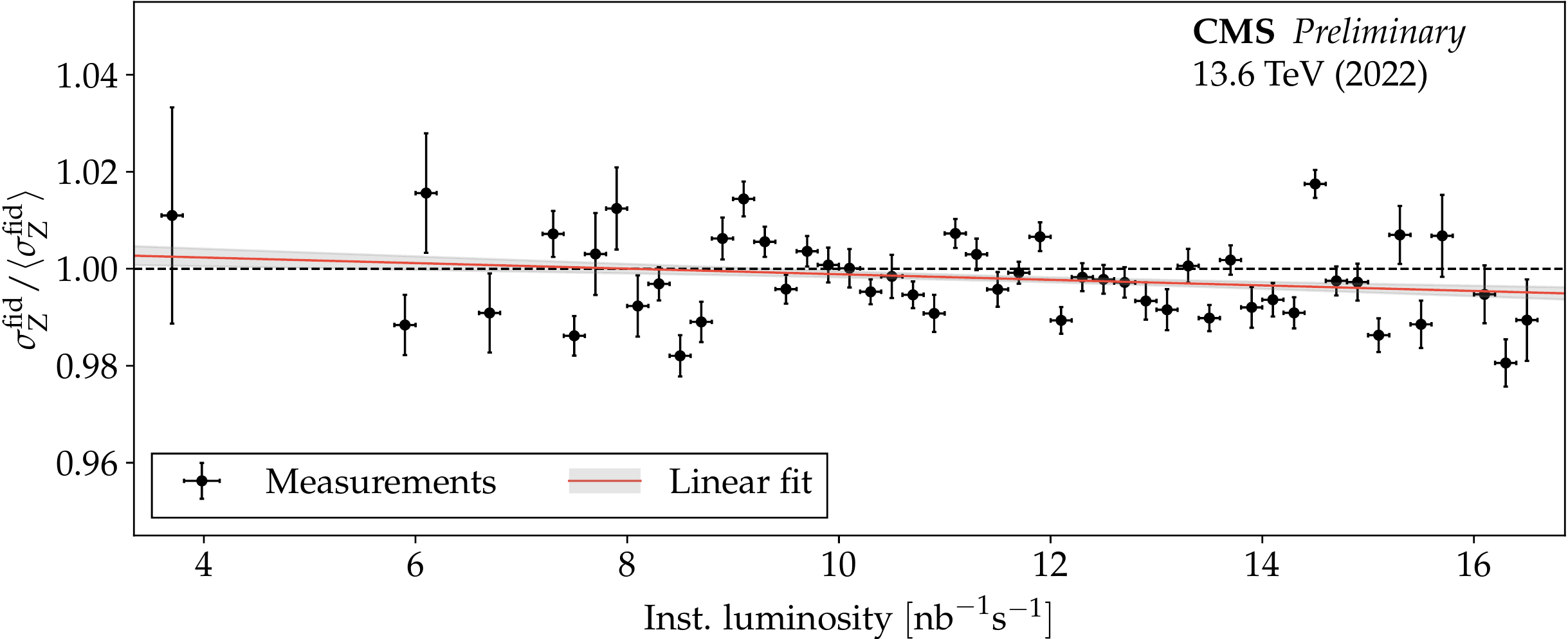}
\caption{%
    (Upper) Integrated luminosity ratio between different luminosity detectors, evaluated in intervals of about 20\,min, as function of total integrated luminosity, for the first 9.6\fbinv\ collected in 2022~\protect\cite{CMS:DP-2022-038}.
    (Lower) Fiducial Z~boson cross section as function of instantaneous luminosity, for the data collected in July and August 2022~\protect\cite{CMS:DP-2023-003}.
}
\label{fig:luminosity}
\end{figure}

The measurement of the CMS luminosity is performed with various subdetectors~\cite{CMS:LUM-17-003}, of which the Pixel Luminosity Telescope~\cite{CMS:NOTE-2022-007,CMS:Karunarathna-2022} and the Fast Beam Conditions Monitor~\cite{CMS:Wanczyk-2022} have been rebuilt and installed before the start of Run 3.
An initial calibration was performed with so-called emittance scans, short beam-separation scans performed in most fills under regular physics data-taking conditions, and good agreement between the measurements provided by the different subdetectors is found~\cite{CMS:DP-2022-038}, as shown in Fig.~\ref{fig:luminosity} (upper).
A full program of Van der Meer beam-separation scans has been performed in November 2022, and interim analysis results are used to improve the calibration of the integrated luminosity measurement.

An independent validation of the luminosity measurement is performed via the counting of Z~boson events~\cite{SalfeldNebgen:Zcounting-2018}.
Dimuon events are selected in intervals of 20\pbinv, and the tag-and-probe method is applied to perform an in~situ evaluation of the HLT, reconstruction, and identification efficiencies of the muons.
Because of the in~situ calibration, the obtained Z~boson counts are independent of changing detector conditions and can be used to evaluate the evolution of the integrated luminosity.
Recently, the CMS Collaboration has demonstrated that the low-pileup data set collected in 2017 can be used to derive a precise calibration of the Z~boson counts, rendering Z~boson counting a fully independent luminosity measurement method~\cite{CMS:LUM-21-001}.
Given that no such data set has been recorded yet at 13.6\,TeV, the fiducial Z~boson cross section (evaluated by normalizing the measured Z~boson counts in each time interval with the measured integrated luminosity) is used to cross-check the linearity and time-stability of the luminosity measurement described above, showing good agreement~\cite{CMS:DP-2023-003}.
The comparison as function of instantaneous luminosity is shown in Fig.~\ref{fig:luminosity} (lower).

\section{Measurement of the top quark pair production cross section}

\begin{figure}[t!]
\centering
\includegraphics[width=0.8\textwidth]{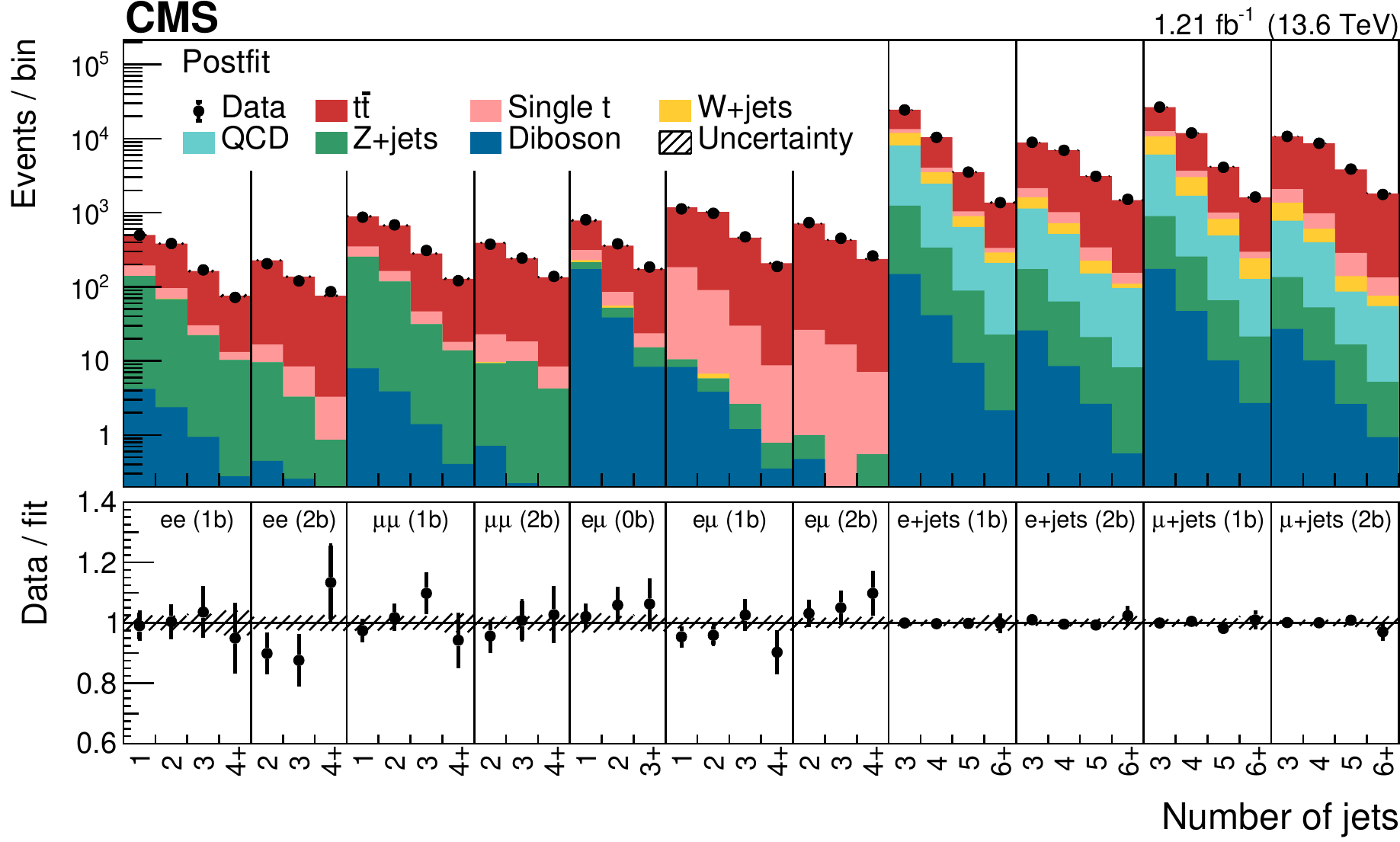}
\caption{%
    Comparison of the number of observed and predicted events in the event categories of the \ttbar\ measurement~\protect\cite{CMS:TOP-22-012}.
    The predictions are shown after fitting the model to the data.
}
\label{fig:ttbar}
\end{figure}

For the first measurement of the top quark pair (\ttbar) production cross section at the new energy~\cite{CMS:TOP-22-012}, a data set collected between 27 July and 03 August 2022 corresponding to an integrated luminosity of 1.21\fbinv\ is analyzed.
Events are selected in the dilepton (e\textmu, ee, \textmu\textmu) and lepton+jets (e+jets, \textmu+jets) channels.
In the e\textmu\ channel, at least one jet is required, and both events with and without b~jets are analyzed.
In the ee and \textmu\textmu\ channels, at least one b~jet is required, and events with dilepton invariant mass consistent with the Z boson mass are removed.
The removed events are then used to validate the Monte Carlo background prediction of Z+jets production.
In the lepton+jets channels, at least three jets are required of which one is required to be a b~jet.
The background from QCD multijet events in the lepton+jets channels is estimated from data sidebands with nonisolated leptons and/or with exactly one jet.
Event categories are formed based on lepton number and flavor, as well as jet and b~jet multiplicity, as shown in Fig.~\ref{fig:ttbar}.

A binned profile likelihood fit simultaneously to all event categories is performed to measure the \ttbar\ cross section.
Lepton identification efficiencies have been estimated with the tag-and-probe method in Z+jets events, and are strongly constrained in the fit from the inclusion of the different lepton flavor combinations.
The b~jet identification efficiencies have not been evaluated prior to the fit but are treated as free parameters, and get constrained in the fit from the inclusion of the different b~jet multiplicity bins.
Besides the integrated luminosity uncertainty of 2.3\%, which is not included in the fit, the lepton and b~jet identification efficiencies are the largest contributions to the systematic uncertainty.
The \ttbar\ cross section is measured to be $882\pm23\,\mathrm{(stat+syst)}\pm20\,\mathrm{(lumi)}\,\mathrm{pb}$, in agreement with the standard model prediction~\cite{Czakon:Ttbar-2014} of $921\,^{+29}_{-37}\,\mathrm{(scale+PDF)}\,\mathrm{pb}$.

To validate the detector performance, several cross checks of the systematic uncertainties are performed using the selected \ttbar\ data set.
The lepton identification efficiencies are integrated over the lepton \pt\ and \abseta, and estimated as free parameters in the fit, yielding results with an uncertainty of about 2\% that are consistent with the dedicated calibration.
The jet energy calibration~\cite{CMS:DP-2022-054} is validated by reconstructing the hadronically decaying W boson in lepton+jets events and fitting its mass, confirming the calibrated jet energy scale within 1.5\%.
Together with the estimation of the b~jet identification efficiencies from data, these results demonstrate that the careful setup of the \ttbar\ analysis with the inclusion of multiple decay channels allows for an in~situ estimation of the most relevant calibrations.

\section{Summary}

\begin{figure}[t!]
\centering
\raisebox{3.5pt}{\includegraphics[width=0.477\textwidth]{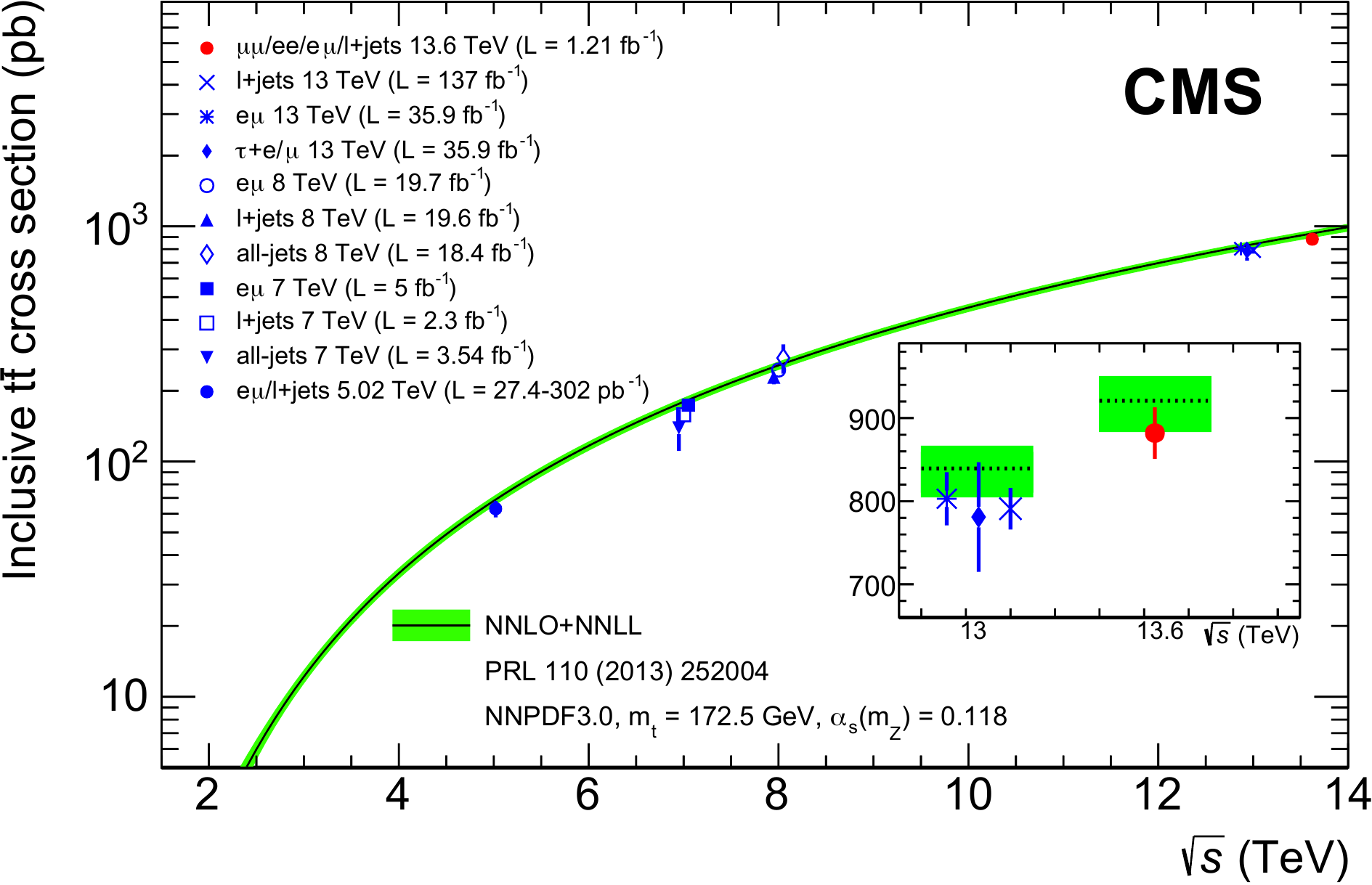}}%
\hspace*{0.05\textwidth}%
\includegraphics[width=0.45\textwidth]{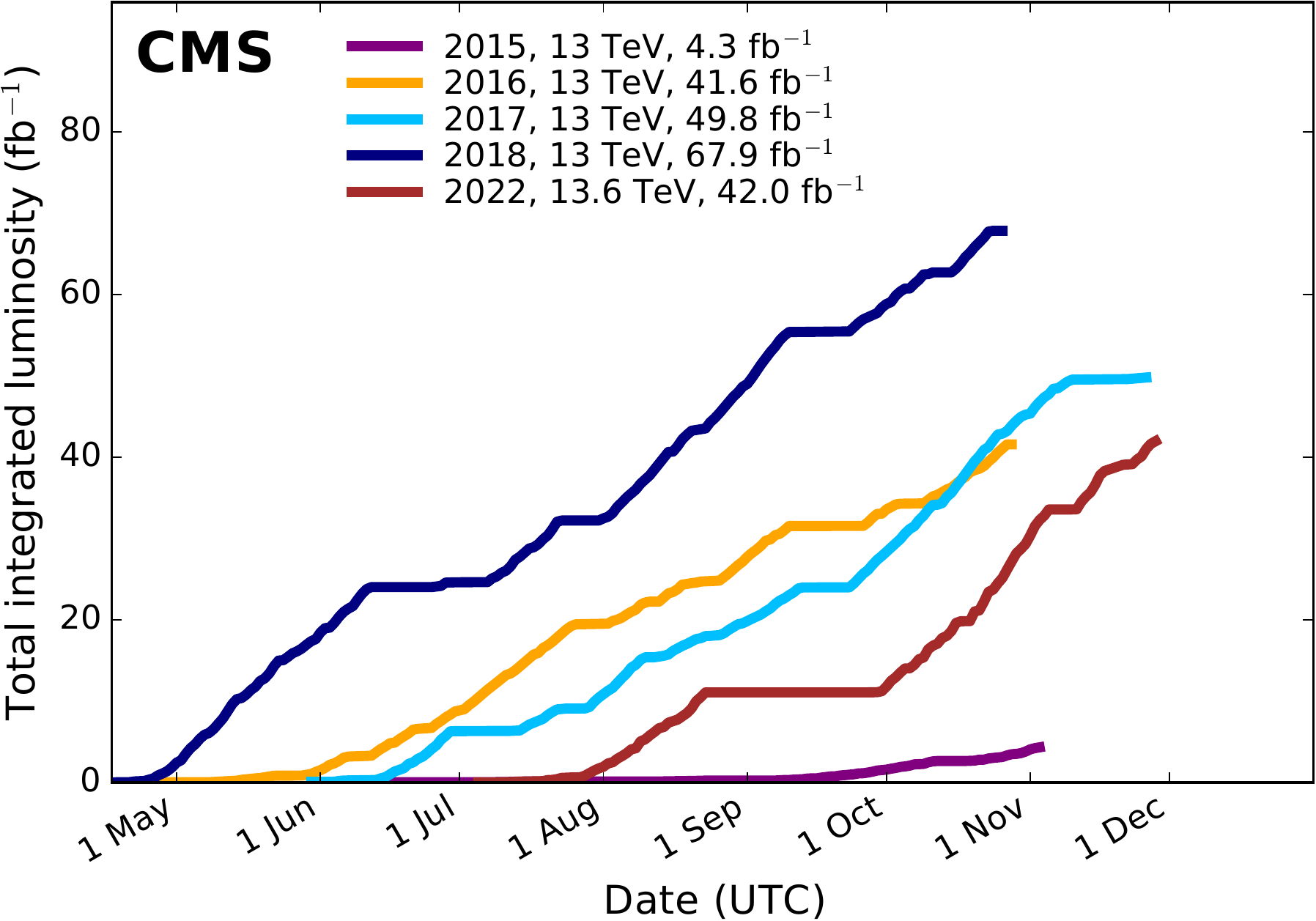}
\caption{%
    (Left) Measured \ttbar\ cross section as function of center-of-mass energy~\protect\cite{CMS:TOP-22-012}.
    (Right) Total integrated luminosity as function of day in the year, separately for each of the 2015--2018 and 2022 data-taking years~\protect\cite{CMS:PublicLumiPlots}.
}
\label{fig:summary}
\end{figure}

The CMS experiment has successfully recorded 38.5\fbinv\ of proton-proton collision data at a center-of-mass energy of 13.6\,TeV in 2022, the first year of Run~3 of the CERN LHC.
The detector performance and object reconstruction is well understood, including the identification of electrons in the trigger, the track reconstruction for muons, and the jet energy calibration.
The luminosity is measured with various CMS subdetectors and cross-checked with a method based on the counting of Z~boson events, showing good agreement between the independent measurements.
A first cross section measurement of \ttbar\ production at 13.6\,TeV has been performed, using 1.21\fbinv\ of data and using both single-lepton and dilepton events.
Comparisons of the measured \ttbar\ cross section and the recorded luminosity in 2022 with previous CMS data sets are shown in Fig.~\ref{fig:summary}.

\section*{References}

\newcommand{\Journal}[6][\relax]{\href{https://doi.org/#6}{\textit{#2}#1 \textbf{#3}, #4 (#5)}}
\newcommand{\Arxiv}[3][submitted to]{\href{http://arxiv.org/abs/#2}{\texttt{arXiv:#2}} (#1 \textit{#3})}
\newcommand{\Cds}[2]{\href{https://cds.cern.ch/record/#2}{#1}}

\end{document}